\documentstyle[preprint,eqsecnum,aps,epsf,tighten]{revtex}

\begin{document}

\newcommand{\dfrac}[2]{\frac{\displaystyle #1}{\displaystyle #2}}
\renewcommand{\mathrm}[1]{{\rm #1}}

\draft
\preprint{VPI--IPPAP--99--09}

\title{Constraints on R--parity violating couplings from lepton universality}
\author{
Oleg~Lebedev\thanks{electronic address: lebedev@quasar.phys.vt.edu},
Will~Loinaz\thanks{electronic address: loinaz@alumni.princeton.edu}, and
Tatsu~Takeuchi\thanks{electronic address: takeuchi@vt.edu}
}
\address{Institute for Particle Physics and Astrophysics,
Physics Department, Virginia Tech, Blacksburg, VA 24061}

\date{Revised : December 7, 1999}
\maketitle

\begin{abstract}
We analyze the one loop corrections to leptonic $W$ and $Z$ decays
in an R--parity violating extension to the Minimal 
Supersymmetric Standard Model (MSSM).  
We find that lepton universality violation 
in the $Z$ line--shape variables alone would strengthen the
bounds on the magnitudes of the $\lambda^\prime$ couplings,
but a global fit on all data leaves the bounds virtually unchanged
at $|\lambda^\prime_{33k}| \leq  0.42$ and 
$|\lambda^\prime_{23k}| \leq 0.50$ at the $2 \sigma$ level.
Bounds from $W$ decays are less stringent:
$|\lambda^\prime_{33k}| \leq 2.4$ at $2 \sigma$, 
as a consequence of the weaker Fermilab experimental bounds
on lepton universality violation in $W$ decays.
We also point out the potential of constraining R--parity
violating couplings from the measurement of the
$\Upsilon$ invisible width.
\end{abstract}


\narrowtext

\section{Introduction}

The assumption of R--parity conservation in supersymmetric model-building 
has long been an economical means of 
(1) avoiding certain phenomenological 
problems in SUSY models ({\it e.g.} 
proton decay), (2) ensuring that the lightest supersymmetric particle 
is available as a cure for the dark matter problem, 
and (3) reducing the SUSY model parameter space.
(For recent reviews, see Ref.~\cite{Dreiner:1997uz}.)
However, the recent discovery of neutrino mass at 
Super--Kamiokande \cite{SuperK:98}
provides improved motivation for R--parity violating extensions to the 
Minimal Supersymmetric Standard Model (MSSM).
Detailed analyses of the phenomenological 
constraints on such models are thus warranted
to quantify the amount of R--parity violation permitted by 
current experimental data.

In this paper we consider the effects of R--parity violating extensions 
to the MSSM on {\it lepton universality} in $W$ and $Z$ decays.  
The R--conserving sector of the MSSM generates lepton universality violations
proportional either to the lepton Yukawa couplings (due to Higgs interactions) 
or to the mass splittings of the sleptons (due to gauge interactions).  
Effects due to the Higgs sector will be considered 
in a future work \cite{LEBEDEV:99}
but will in general be negligible unless $\tan\beta$ is quite large
\cite{HISANO:97}.
Effects due to gauge interactions are negligible if the slepton mass 
splittings are small.  
This is the case, for example, 
in supergravity (SUGRA) models with universal soft--breaking scalar 
masses at the SUGRA scale, in which the mass degeneracy is broken 
only by renormalization group running effects involving small 
Yukawa couplings.  In R--parity violating models, however, 
R--parity violating interactions provide additional sources of 
lepton universality violation which may be significantly larger than
these smaller effects.

The R--parity violating superpotential has the following
form:\footnote{Because we are neglecting the soft breaking terms, we can rotate the bilinear 
terms 
away \cite{Hall:1984id}.
For possible effects of the soft breaking terms on $W$ and $Z$ decays
see Ref.~\cite{NOWAKOWSKI:96}.}
\begin{equation}
W_{\not{R}}
= \frac{1}{2} \lambda_{ijk}   \hat{L}_i \hat{L}_j \hat{E}_k 
+             \lambda_{ijk}'  \hat{L}_i \hat{Q}_j \hat{D}_k 
+ \frac{1}{2} \lambda_{ijk}'' \hat{U}_i \hat{D}_j \hat{D}_k\;,
\label{eq:superpotential}
\end{equation}
where $\hat{L}_i$, $\hat{E}_i$, $\hat{Q}_i$, $\hat{U}_i$, and 
$\hat{D}_i$ are
the MSSM superfields defined in the usual fashion \cite{Haber:1985rc}, 
and the subscript $i=1,2,3$ is the generation index.  
Since {\it a priori} the interactions described by this superpotential have
an arbitrary flavor structure, we generically expect that in the context of 
$W$ and $Z$ decays these will give rise to lepton universality violations.  
In this paper, we estimate the size of this violation
and derive constraints on the R--parity violating couplings from LEP 
and Fermilab measurements of the leptonic observables in $W$ and $Z$ decays. 

It is clear that the purely baryonic operator 
$\hat{U}_i \hat{D}_j \hat{D}_k$ is irrelevant to our discussion.  
The other two operators may affect W and Z decays at one loop through vertex 
corrections to the $W\ell_L\bar{\nu}$ and $Z\ell_L \bar{\ell}_L$ vertices, 
with superparticles running in the loop.  
However, the couplings $\lambda_{ijk}$ are already tightly constrained 
to be at most ${\cal O}(10^{-2})$,
as the operator $\hat{L}_i \hat{L}_j \hat{E}_k$ violates 
lepton universality in lepton decays at tree level \cite{Barger:1989rk}.  
The constraints on $\lambda_{ijk}'$ are much less stringent.  
Previous limits cite upper bounds on $\lambda_{ijk}'$ as large as the 
gauge couplings ({\it i.e.} as large as 0.5) with the SUSY 
scale at 100 GeV \cite{Dreiner:1997uz}.
One may thus expect significant radiative corrections 
induced by these couplings.
Henceforth we will focus on the effects of the operator
$\hat L_{i} \hat Q_{j} \hat D_{k}$ only.

It is important to note that very strict constraints on the 
products of {\em different} R--violating couplings already exist
from flavor-changing processes, {\it e.g.} 
$\mu\rightarrow e\gamma$ constrains 
$|\lambda^{\prime}_{1ij}\lambda^{\prime}_{2ij}| < 4.6 \times 10^{-4}$ 
\cite{deCarlos:1996du}.  
However, these constraints can be easily satisfied by requiring
only one of these couplings to be very small leaving the other
coupling ill--constrained. 
The $W$ and $Z$ decay processes we consider here are 
flavor--conserving and involve the {\em same} R--violating coupling squared.  
Thus we can constrain the individual couplings rather than their products.  
 
We emphasize that, in the absence of a complete calculation 
in the full theory, focussing attention on the violation of
lepton universality provides a clear advantage over
studying the effects of R--breaking couplings 
on the individual lepton--gauge boson couplings separately.
This is because the R--conserving sector induces significant
{\it universal corrections} to the lepton couplings which 
depend strongly on the choice of SUSY parameters.  
These corrections (along with the corrections to the hadronic 
partial widths) cancel when considering violations of 
lepton universality.
Thus, the study of lepton universality violation lets us
isolate the effects of R--breaking interactions without
ad hoc assumptions about corrections from the 
R--conserving sector.

In the following calculations we neglect left--right squark mixing. 
Left--right squark mixing could be large only for the stop.  
However, since diagrams involving the stop contain down quarks with 
negligible mass, it will be seen that the contributions from these 
diagrams are numerically small
(subleading in an expansion in $m_W^2$ or $m_Z^2$).  
Further, due to the chiral structure of the R--breaking interactions, 
two left--right mass insertions would be required in the diagram.  
Thus, such contributions would be further suppressed as long as 
the mixing parameter is perturbatively small.

Radiative corrections to individual $Z\rightarrow \ell\bar{\ell}$ 
partial widths due to R--breaking interactions have previously been 
considered in Ref.~\cite{Bhattacharyya:1995pr} but lacked a consistent
treatment of the R--conserving corrections.
In this paper, we study the violation of 
lepton universality in $W$ and $Z$ decay to isolate the effects 
of R--breaking couplings and constrain their sizes.
In determining the limits on R--breaking from $Z$ decay, 
we perform a global fit to all the relevant LEP and
SLD observables in which
the corrections from both R--breaking and R--conserving interactions
are parametrized and fit to the data.
This provides a consistent accounting of R--conserving effects and
allows us to improve the existing bounds on the 
R--breaking $\lambda^{\prime}$ couplings.  A companion study of constraints
on $\lambda'$ and $\lambda''$ couplings from LEP/SLD hadronic observables
has been performed in Ref.~\cite{Lebedev:1999ze}.

\section{Leptonic W decays}
\label{sec:Wdecay}

\begin{figure}[ht]
\begin{center}
\unitlength=1cm
\begin{picture}(13,6)(0,0)
\unitlength=1mm
\put(21,27){\vector(1,0){8}}
\put(23,21){$Q$}
\put(48,42){\vector(2,1){6}}
\put(48,44){$p$}
\put(48,18){\vector(2,-1){6}}
\put(48,14){$q$}
\put(40,2){(a)}
\put(96,2){(b)}
\put(14,29){$W^-$}
\put(69,29){$W^-$}
\put(59,44){$e_{i_L}$}
\put(114,44){$e_{i_L}$}
\put(59,16){$\nu_{i'_L}$}
\put(114,16){$\nu_{i'_L}$}
\put(50,29){$d_{j_R}$}
\put(105,29){$\tilde{d}_{k_R}$}
\put(36,39){$\tilde{u}_{k_L}$}
\put(36,20){$\tilde{d}_{k_L}$}
\put(92,39){$u_{j_L}$}
\put(92,21){$d_{j_L}$}
\epsfbox[0 0 360 150]{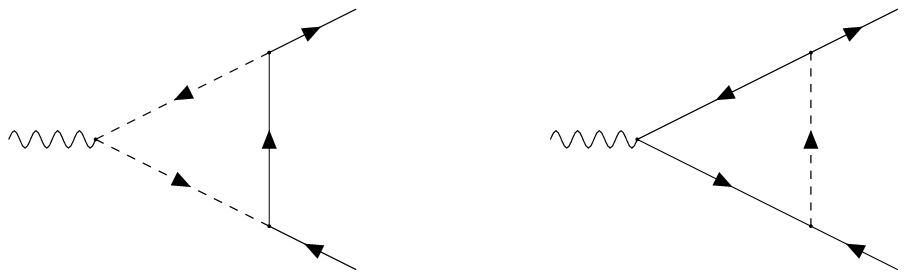}
\end{picture}
\caption{Vertex corrections to $W^-\rightarrow e_{i_L}\bar{\nu}_{i'_L}$
from R--parity violating interactions.}
\end{center}
\label{Wdecay1}
\end{figure}

\begin{figure}[ht]
\begin{center}
\unitlength=1cm
\begin{picture}(13,11)(0,0)
\unitlength=1mm
\put(40,55){(a)}
\put(96,55){(b)}
\put(40,2){(c)}
\put(96,2){(d)}
\put(14,30){$W^-$}
\put(69,30){$W^-$}
\put(14,79){$W^-$}
\put(69,79){$W^-$}
\put(59,44){$e_{i_L}$}
\put(114,44){$e_{i_L}$}
\put(59,16){$\nu_{i'_L}$}
\put(114,16){$\nu_{i'_L}$}
\put(59,93){$e_{i_L}$}
\put(114,93){$e_{i_L}$}
\put(59,65){$\nu_{i'_L}$}
\put(114,65){$\nu_{i'_L}$}
\put(44,82){$d_{k_R}$}
\put(35,95){$\tilde{u}_{j_L}$}
\put(30,86){$e_{i'_L}$}
\put(100,76){$d_{k_R}$}
\put(85,74){$\nu_{i_L}$}
\put(92,63){$\tilde{d}_{j_L}$}
\put(40,42){$u_{j_L}$}
\put(48,29){$\tilde{d}_{k_R}$}
\put(28,36){$e_{i'_L}$}
\put(85,25){$\nu_{i_L}$}
\put(95,20){$d_{j_L}$}
\put(104,30){$\tilde{d}_{k_R}$}
\epsfbox[0 470 360 780]{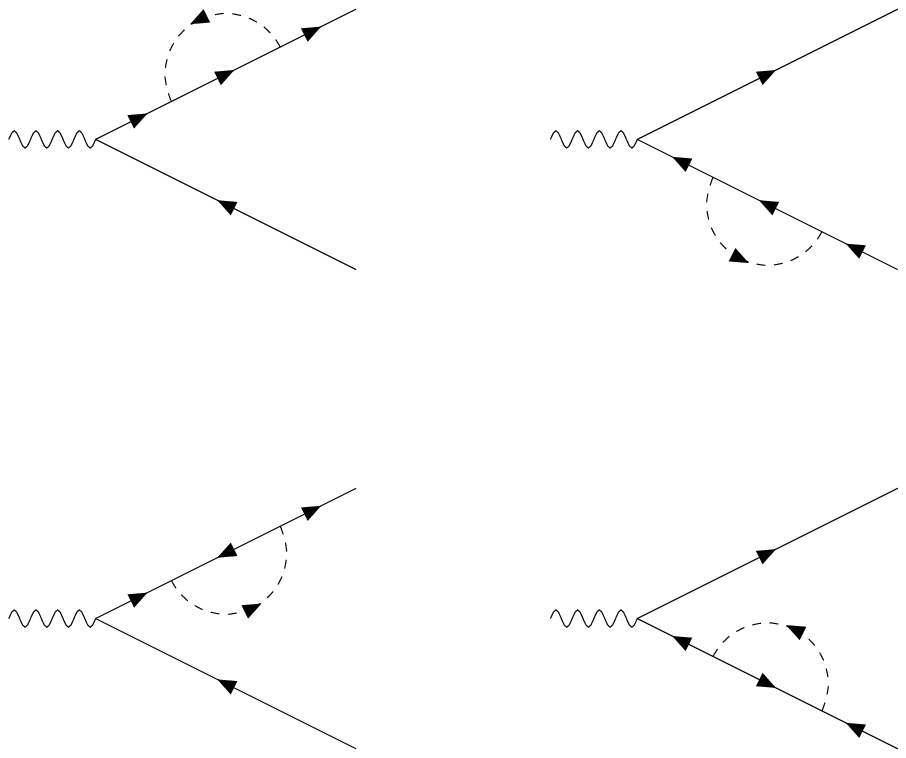}
\end{picture}
\caption{Wavefunction renormalization corrections to
$W^-\rightarrow e_{i_L}\bar{\nu}_{i'_L}$ from R--parity violating
interactions.}
\end{center}
\label{Wdecay2}
\end{figure}

The relevant R--parity violating interactions expressed in terms of 
the component fields take the form
\begin{eqnarray}
\lefteqn{
\Delta{\cal L}_{\not{R}}
    = \lambda'_{ijk}
      \bigg[ \tilde{\nu}_{iL} \overline{d}_{kR} d_{jL}
            + \tilde{d}_{jL} \overline{d}_{kR} \nu_{iL}
            + \tilde{d}^*_{kR} \overline{\nu}^c_{iL} d_{jL}
} \cr
& &
            - (\tilde{e}_{iL} \overline{d}_{kR} u_{jL}
            + \tilde{u}_{jL} \overline{d}_{kR} e_{iL}
            + \tilde{d}^*_{kR} \overline{e}^c_{iL} u_{jL})
      \bigg]\; + \;{h.c.}
\label{eq:Lagrangian}
\end{eqnarray}
The one loop diagrams contributing to the decay
$W \rightarrow e_{iL} \bar{\nu}_{i^\prime L}$ 
are shown in Figs. \ref{Wdecay1} and \ref{Wdecay2}.  
At one loop, the neutrino flavor may differ from that of the 
tree level vertex ($i\neq i'$) as a result of the 
R--parity violating interactions.
Since neutrino flavor is indistinguishable in the detector, 
we should in principle sum over all three generations of 
antineutrino in the final state.  
However, since this is a one loop effect which does not 
interfere with the tree level flavor
conserving decay, we will neglect it in our analysis and
set $i=i'$.

The amplitude of each diagram in Figs.~\ref{Wdecay1} and
\ref{Wdecay2} are:
\begin{eqnarray}
\lefteqn{%
-N_C\, | \lambda^\prime_{ijk} |^2\,
\Biggl[ -i\frac{ g }{ \sqrt{2} }\,
        W^\mu(p+q)\,\bar{e}_{iL}(p) \gamma_\mu \nu_{iL}(q)\,
\Biggr] \times}
\qquad & & \cr
(1a) &:& 2\,\hat{C}_{24}
          \!\left( 0,0,m_W^2;0,m_{\tilde{u}_{jL}},m_{\tilde{d}_{jL}}
            \right) \cr
(1b) &:& (d-2)\,\hat{C}_{24}
          \!\left( 0,0,m_W^2;m_{\tilde{d}_{kR}},m_{u_j},0
            \right) \cr
     & &
       - m_W^2\,\hat{C}_{23}\!\left( 0,0,m_W^2;m_{\tilde{d}_{kR}},m_{u_j},0
                              \right) \cr
(2a) &:& B_1\!\left( 0;0,m_{\tilde{u}_{jL}} \right) \cr
(2b) &:& B_1\!\left( 0;0,m_{\tilde{d}_{jL}} \right) \cr
(2c) &:& B_1\!\left( 0;m_{u_j},m_{\tilde{d}_{kR}} \right) \cr
(2d) &:& B_1\!\left( 0;      0,m_{\tilde{d}_{kR}} \right)
\label{eq:Wamplitude}
\end{eqnarray}
The expression in the square brackets is the tree level amplitude.
The definitions of the integrals $B_1$, $\hat{C}_{23}$, and $\hat{C}_{24}$ 
are presented in the Appendix.
In the above expressions $N_C=3$ is the number of colors,
and $i$, $j$, $k$ are family indices;
the final result must be summed over $j$ and $k$ to obtain
the full correction for final state flavor $i$.
We have set all the down type quark masses to zero.
The up type quark masses $m_{u_j}$ will also be set to zero
except for the top quark ($j=3$, $m_{u_3} = m_t$).

Combining the expressions in Eq.~(\ref{eq:Wamplitude}),
with appropriate factors of $\frac{1}{2}$ for the wavefunction
renormalizations, we obtain the one--loop shift of the 
$W e_{iL}\bar{\nu}_{iL}$ coupling due to the $\lambda'_{ijk}$
interaction:
\widetext
\begin{eqnarray}
\delta g_{ijk}
& = & \delta g_{ijk}^{(u)} + \delta g_{ijk}^{(d)}, \cr
&   & \cr
\frac{ \delta g_{ijk}^{(d)} }{ g }
& \equiv & - N_C | \lambda^{\prime}_{ijk} |^2
      \Bigg[ 2\,\hat{C}_{24}\left( 0,m_{\tilde{u}_{jL}},m_{\tilde{d}_{jL}}
                             \right)
            + \frac{1}{2} B_1\left( 0,m_{\tilde{u}_{jL}} \right)
            + \frac{1}{2} B_1\left( 0,m_{\tilde{d}_{jL}} \right)
      \Bigg],
\cr
\frac{ \delta g_{ijk}^{(u)} }{ g }
& \equiv & - N_C | \lambda^{\prime}_{ijk} |^2
      \Bigg[ (d-2) \hat{C}_{24}\left( m_{\tilde{d}_{kR}},m_{u_j},0
                               \right)  
           - m_W^2 \hat{C}_{23}\left( m_{\tilde{d}_{kR}},m_{u_j},0
                               \right) 
\cr
&   & \qquad\qquad\qquad\qquad\qquad\qquad\quad
           + \frac{1}{2} B_1\left(       0,m_{\tilde{d}_{kR}} \right)
           + \frac{1}{2} B_1\left( m_{u_j},m_{\tilde{d}_{kR}} \right)
      \Bigg].
\end{eqnarray}
\narrowtext
We have suppressed the external momentum dependence of the $B$ and $C$
functions to simplify our expressions.
The contributions of diagrams involving the down--type quark
have been combined in $\delta g_{ijk}^{(d)}$ and those 
involving the up--type quark in $\delta g_{ijk}^{(u)}$.  
The $1/\epsilon$ poles
of dimensional regularization cancel separately in each of these
combinations so they are finite.

In the following, we evaluate the size of these shifts
for a common squark mass of $m_{\tilde{q}}=100$~GeV.
To facilitate estimations for different squark masses, we provide
approximate formulae.

\begin{itemize}

\item
First, we evaluate $\delta g_{i3k}^{(u)}$, the contribution from
diagrams involving the top quark $u_3$.
For $m_{u_3} = m_t = 175$~GeV,
$m_W = 81$~GeV, and $m_{\tilde{q}} = 100$~GeV, we find:
\begin{equation}
\frac{ \delta g_{i3k}^{(u)} }{ g }
= -1.02\%\, | \lambda^\prime_{i3k} |^2
\end{equation}
An approximate expression can be obtained by expanding
the full expression of $\delta g_{i3k}^{(u)}$ in powers of $m_W^2$:
\begin{eqnarray}
\frac{ \delta g_{i3k}^{(u)} }{ g }
& \approx & -\frac{N_C}{(4 \pi)^2} |\lambda^{\prime}_{i3k}|^2 \cr
&   & \times\bigg[\,\frac{ x }{ 4 ( 1 - x )^2 }
            \Big\{\, x - 1 - ( 2 - x ) \ln{ x } \,
            \Big\}  \cr
&   & \phantom{\times}
         +  \frac{m_W^2}{m_t^2}\,\frac{ x }{ 3 ( 1 - x )^2 }
            \Big\{ 1 - x + \ln{x}
            \Big\}
      \bigg]
\label{eq:Wleading}
\end{eqnarray}
where $x = m_t^2/m^2_{\tilde{q}}$.
For $m_{\tilde{q}} = 100$~GeV, this expression is equal to
$(-1.11+0.09)\%\, |\lambda^{\prime}_{i3k}|^2$.
Compared to the exact result above, we see that the leading
order approximation is already fairly accurate.

\item
The contributions of the diagrams with massless quarks are
numerically smaller and vanish as $m_W\rightarrow 0$.
The correction with massless up-type quarks ($j=1,2$) is:
\begin{equation}
\frac{ \delta g_{ijk}^{(u)} }{ g }
= 0.22\%\, | \lambda^\prime_{ijk} |^2\qquad (j=1,2)
\end{equation}
The approximate form to leading order in $m_W^2/m_{\tilde{q}}^2$ is
\begin{equation}
\frac{ \delta g_{ijk}^{(u)} }{ g }
\approx \frac{ N_C }{ (4\pi)^2 } |\lambda^{\prime}_{ijk}|^2\;  
        \frac{ m_W^2 }{ 9 m^2_{\tilde{q}} } 
        \left( 1 - 3\ln\frac{ m_W^2 }{ m^2_{\tilde{q}} }
        \right).
\label{eq:Wsubleading}
\end{equation}
For $m_{\tilde{q}} = 100$~GeV, this gives
$0.31\%\, |\lambda^{\prime}_{ijk}|^2$ which suffices
for our purpose.

\item
The diagrams with down--type quarks contribute:
\begin{equation}
\frac{ \delta g_{ijk}^{(d)} }{ g }
= 0.07\%\, | \lambda^\prime_{ijk} |^2
\end{equation}
the approximate expression being
\begin{equation}
\frac{ \delta g_{ijk}^{(d)} }{ g }
\approx \frac{N_C}{(4 \pi)^2} |\lambda^{\prime}_{ijk}|^2 
      \left( \frac{ m_W^2 }{ 18\,m^2_{\tilde{q}} }
      \right).
\label{eq:Wsubsubleading}
\end{equation}
For $m_{\tilde{q}} = 100$~GeV, this gives
$0.07\%\, |\lambda^{\prime}_{ijk}|^2$ to the accuracy shown.

\end{itemize}
Note that each contribution, Eqns.~(\ref{eq:Wleading}), (\ref{eq:Wsubleading}),
and (\ref{eq:Wsubsubleading}) separately decouples in the limit
$m_{{\tilde q}}^2 \rightarrow \infty$ as they should.
Collecting everything together, the shift in the coupling of the $i$--th
generation lepton to the $W$ is given by:
\widetext
\begin{equation}
\frac{ \delta g_i }{ g }
= \sum_{j,k}\left[ \frac{ \delta g_{ijk}^{(u)} }{ g }
                 + \frac{ \delta g_{ijk}^{(d)} }{ g }
            \right]
= -0.95\% \sum_{k} | \lambda'_{i3k} |^2
  +0.29\% \sum_{k} | \lambda'_{i2k} |^2
  +0.29\% \sum_{k} | \lambda'_{i1k} |^2
\end{equation}
\narrowtext
where we have summed over all possible generation indices $j$ and $k$.

The current bound on lepton universality violation in leptonic 
$W$ decays from D{0\kern-6pt/} is \cite{Rimondi:1999pu}
\begin{eqnarray*}
\frac{ g_\tau }{ g_e } = 1.004 \pm 0.019 (stat.) \pm 0.026 (syst.).
\end{eqnarray*}
This places a constraint on
\begin{equation}
\delta\left( \frac{ g_\tau }{ g_e } \right)
= \frac{ \delta g_3 }{ g } - \frac{ \delta g_1 }{ g }.
\end{equation} 
Note that R--conserving corrections do not contribute since
they cancel in the ratio $g_\tau/g_e$.
The constraint on the R--breaking couplings is thus
\begin{eqnarray}
\lefteqn{-0.4 \pm 1.9 (stat.) \pm 2.6 (syst.)}
\rule[-2mm]{0mm}{6mm}\cr
& = & \Big\{ \sum_k |\lambda'_{33k}|^2 
       - 0.3 \sum_k |\lambda'_{32k}|^2
       - 0.3 \sum_k |\lambda'_{31k}|^2
      \Big\} \cr
& - & \Big\{ \sum_k |\lambda'_{13k}|^2
       - 0.3 \sum_k |\lambda'_{12k}|^2
       - 0.3 \sum_k |\lambda'_{11k}|^2
      \Big\}. 
\label{eq:WdecayConstraint}
\end{eqnarray}
Of the couplings $\lambda'_{ijk}$ 
appearing in this expression, the $i=1$ couplings
are already well constrained from neutrino--less
double beta decay, neutrino masses, atomic parity violation, and
low energy charged current universality.
Using the 2$\sigma$ limits for the individual couplings
charted in Ref.~\cite{Allanach:1999ic},
we find for $m_{\tilde{q}}=100$~GeV:
\begin{eqnarray}
\sum_k |\lambda^{\prime}_{11k}|^2 & \leq & 0.00088,\cr
\sum_k |\lambda^{\prime}_{12k}|^2 & \leq & 0.0055, \cr
\sum_k |\lambda^{\prime}_{13k}|^2 & \leq & 0.079.
\label{eq:1jkconstraint}
\end{eqnarray}
The $i=3$, $j=1$ couplings are also well constrained from
$R_{\tau\pi} = \Gamma(\tau\rightarrow\pi\nu_\tau)/
\Gamma(\pi\rightarrow\mu\nu_\mu)$.  Again, using the
2$\sigma$ limits cited in Ref.~\cite{Allanach:1999ic} we find
\begin{equation}
\sum_k |\lambda'_{31k}|^2 \leq 0.036
\label{eq:31kconstraint}
\end{equation}
Therefore, we can neglect the $\lambda'_{1jk}$ and
$\lambda'_{31k}$ terms in Eq.~(\ref{eq:WdecayConstraint})
and obtain
\begin{equation}
\sum_k | \lambda'_{33k} |^2 - 0.3\sum_k | \lambda'_{32k} |^2
= - 0.4 \pm 3.2
\label{eq:WdecayConstraint2}
\end{equation}
where the systematic and statistical errors have been added in quadrature.
If we neglect the $32k$ term with a smaller numerical coefficient,
this places a 1$\sigma$ (2$\sigma$) upper bound on
the $33k$ term:
\begin{equation}
\sum_k | \lambda'_{33k} |^2 \leq 2.8 \;(6.0),
\end{equation}
which in turn translates into the limit
\begin{equation}
|\lambda'_{33k}| \leq 1.7\;(2.4).
\end{equation}
Non--zero values of $\lambda'_{32k}$ will weaken this bound.

As we will see later, $Z$ decay data places a constraint
on $\sum_k |\lambda'_{33k}|^2$ at the $\pm 0.1$ level.
Therefore, for the $W$ decay data to be competitive with
the $Z$ decay data, the error must be improved by more than an
order of magnitude.
While the Tevatron Run II may provide enough data to
improve the statistical error considerably, improving the
systematic error may prove a challenge \cite{Rimondi:Private}.

\section{Leptonic Z decays}

\begin{figure}[ht]
\begin{center}
\unitlength=1cm
\begin{picture}(13,15)(0,0)
\unitlength=1mm
\put(22,127){\vector(1,0){8}}
\put(24,121){$Q$}
\put(49,142){\vector(2,1){6}}
\put(49,144){$p$}
\put(49,117){\vector(2,-1){6}}
\put(49,113){$q$}
\put(40,105){(a)}
\put(96,105){(b)}
\put(40,52){(c)}
\put(96,52){(d)}
\put(68,2){(e)}
\put(16,80){$Z$}
\put(71,80){$Z$}
\put(16,129){$Z$}
\put(71,129){$Z$}
\put(44,31){$Z$}
\put(86,16){$e_{i_L}$}
\put(86,46){$e_{i_L}$}
\put(59,94){$e_{i_L}$}
\put(114,94){$e_{i_L}$}
\put(59,66){$e_{i_L}$}
\put(114,66){$e_{i_L}$}
\put(59,143){$e_{i_L}$}
\put(114,143){$e_{i_L}$}
\put(59,115){$e_{i_L}$}
\put(114,115){$e_{i_L}$}
\put(36,90){$u_{j_L}$}
\put(36,71){$u_{j_L}$}
\put(50,80){$\tilde{d}_{k_R}$}
\put(92,90){$d_{k_R}$}
\put(92,71){$d_{k_R}$}
\put(105,80){$\tilde{u}_{j_L}$}
\put(36,138){$\tilde{u}_{j_L}$}
\put(36,120){$\tilde{u}_{j_L}$}
\put(50,128){$d_{k_R}$}
\put(92,138){$\tilde{d}_{k_R}$}
\put(92,119){$\tilde{d}_{k_R}$}
\put(105,128){$u_{j_L}$}
\put(58,26){$u_{j_R}$}
\put(58,39){$u_{j_R}$}
\put(68,22){$u_{j_L}$}
\put(68,43){$u_{j_L}$}
\put(78,31){$\tilde{d}_{k_R}$}
\epsfbox[0 330 360 780]{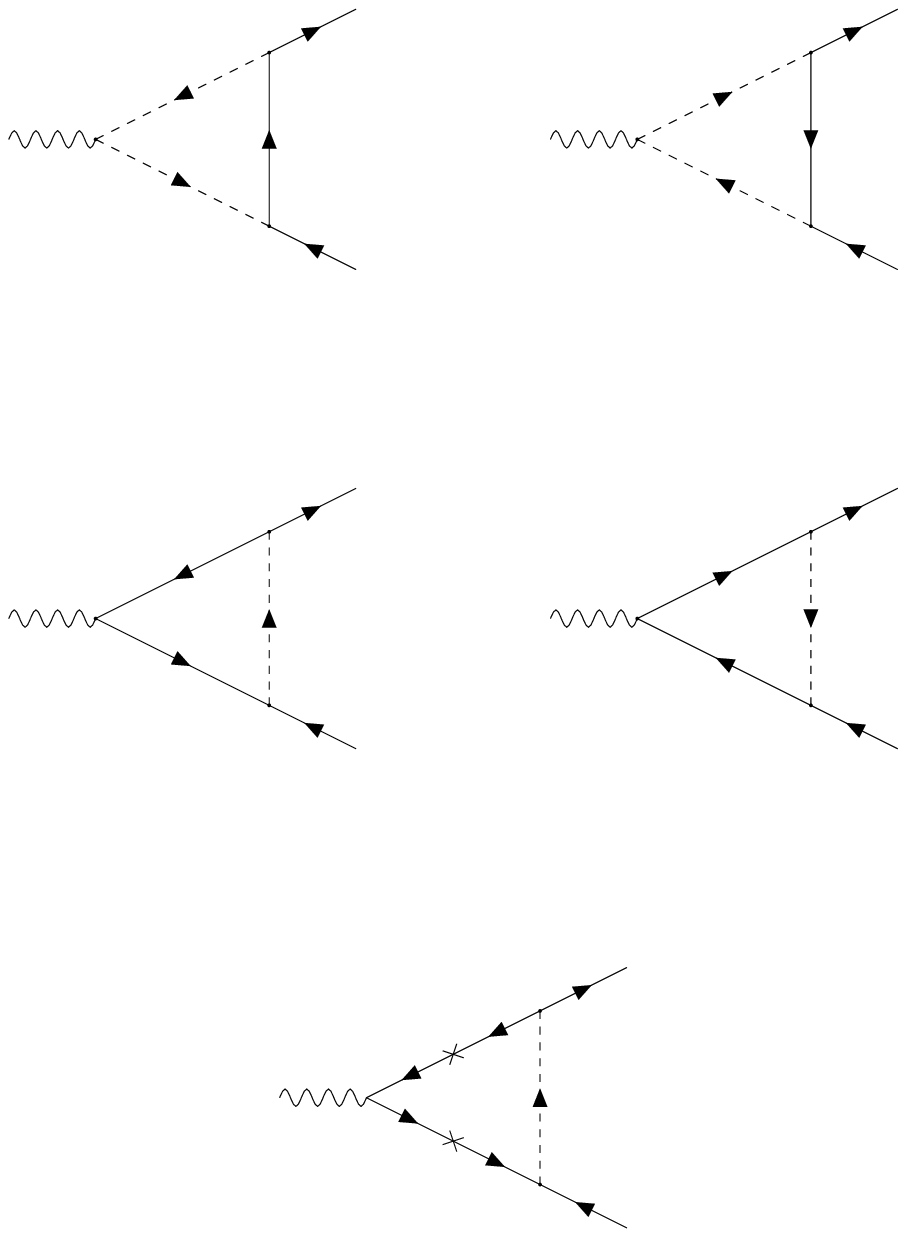}
\end{picture}
\caption{Vertex corrections to $Z\rightarrow e_{i_L}\bar{e}_{i_L}$
from R--parity violating interactions.}
\end{center}
\label{Zdecay1}
\end{figure}

\begin{figure}[ht]
\begin{center}
\unitlength=1cm
\begin{picture}(13,11)(0,0)
\unitlength=1mm
\put(40,55){(a)}
\put(96,55){(b)}
\put(40,2){(c)}
\put(96,2){(d)}
\put(14,30){$Z$}
\put(69,30){$Z$}
\put(14,79){$Z$}
\put(69,79){$Z$}
\put(59,44){$e_{i_L}$}
\put(114,44){$e_{i_L}$}
\put(59,16){$e_{i_L}$}
\put(114,16){$e_{i_L}$}
\put(59,93){$e_{i_L}$}
\put(114,93){$e_{i_L}$}
\put(59,65){$e_{i_L}$}
\put(114,65){$e_{i_L}$}
\put(44,82){$d_{k_R}$}
\put(35,95){$\tilde{u}_{j_L}$}
\put(30,86){$e_{i_L}$}
\put(100,76){$d_{k_R}$}
\put(85,74){$e_{i_L}$}
\put(92,63){$\tilde{u}_{j_L}$}
\put(40,42){$u_{j_L}$}
\put(48,29){$\tilde{d}_{k_R}$}
\put(28,36){$e_{i_L}$}
\put(85,25){$e_{i_L}$}
\put(95,20){$u_{j_L}$}
\put(104,30){$\tilde{d}_{k_R}$}
\epsfbox[0 470 360 780]{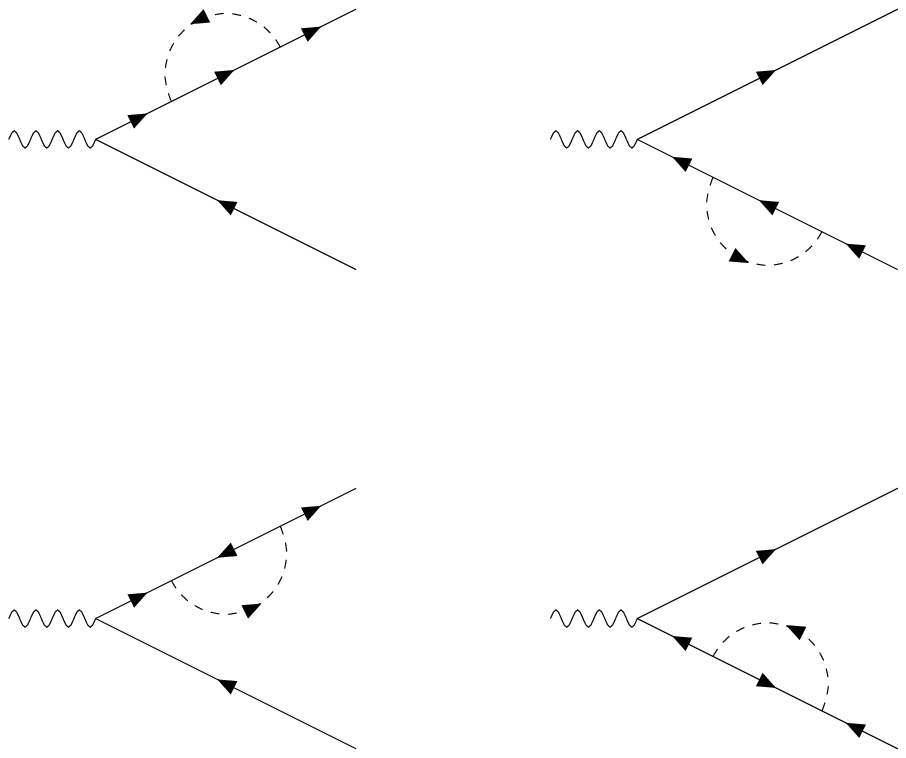}
\end{picture}
\caption{Wavefunction renormalization corrections to
$Z\rightarrow e_{i_L}\bar{e}_{i_L}$
from R--parity violating interactions.}
\end{center}
\label{Zdecay2}
\end{figure}

We next consider the effect of R--parity violating interactions on 
flavor--conserving leptonic $Z$ decays, 
$Z\rightarrow e_{iL} \bar{e}_{iL}$.
Note that the $\lambda'$ interaction in Eq.~(\ref{eq:superpotential})
involves only the left--handed lepton field.
Therefore, at the one--loop level the right--handed coupling is unaffected.
Neglecting all down--type quark masses, the amplitudes of the
diagrams shown in Figs.~\ref{Zdecay1} and \ref{Zdecay2} are
\begin{eqnarray}
\lefteqn{%
-N_C | \lambda^\prime_{ijk} |^2
\Biggl[ -i\frac{ g }{ \cos\theta_W }\,
        Z^\mu(p+q)\,\bar{e}_{iL}(p) \gamma_\mu e_{iL}(q)\,
\Biggr] \times}
\qquad & & \cr
(3a) & : & -2 h_{u_L}
     \hat{C}_{24}\left( 0,0,m_Z^2;0,m_{\tilde{u}_{jL}},m_{\tilde{u}_{jL}}
                 \right) 
\cr
(3b) & : & +2 h_{d_R}
     \hat{C}_{24}\left( 0,0,m_Z^2;m_{u_j},m_{\tilde{d}_{kR}},m_{\tilde{d}_{kR}}
                 \right)
\cr
(3c) & : & -  h_{u_L}
\left[ (d-2)\hat{C}_{24}\left( 0,0,m_Z^2;m_{\tilde{d}_{kR}},m_{u_j},m_{u_j}
                        \right) 
\right. \cr
     &   &
\left.\qquad\quad
    - m_Z^2 \hat{C}_{23}\left( 0,0,m_Z^2;m_{\tilde{d}_{kR}},m_{u_j},m_{u_j}
                        \right) 
\right]
\cr
(3d) & : & +  h_{d_R}
\left[ (d-2)\hat{C}_{24}\left( 0,0,m_Z^2;m_{\tilde{u}_{jL}},0,0
                        \right)  
\right. \cr
     &   &
\left.\qquad\quad
    - m_Z^2 \hat{C}_{23}\left( 0,0,m_Z^2;m_{\tilde{u}_{jL}},0,0
                        \right) 
\right]
\cr
(3e) & : & h_{u_R} m_{u_j}^2  
           \hat{C}_0\left( 0,0,m_Z^2;m_{\tilde{d}_{kL}},m_{u_j},m_{u_j}
                    \right)
\cr
(4a) & + & (4b)\;:\; 2 h_{e_L} 
                  B_1\left( 0;      0,m_{\tilde{u}_{jL}} \right)  \cr
(4c) & + & (4d)\;:\; 2 h_{e_L} 
                  B_1\left( 0;m_{u_j},m_{\tilde{d}_{kR}} \right)
\end{eqnarray}
where
\begin{equation}
h_{f_L} = I_3 - Q_f \sin^2\theta_W, \qquad
h_{f_R} =     - Q_f \sin^2\theta_W.
\end{equation}
The tree level amplitude is $h_{e_L}$ times
the expression in the square brackets.
These corrections can be expressed as a
shift in the coupling $h_{e_L}$:
\widetext
\begin{eqnarray}
\delta h_{ijk} 
& = & \delta h_{ijk}^{(u)} + \delta h_{ijk}^{(d)}  \cr
& & \cr
\delta h_{ijk}^{(d)}
& \equiv & - N_C |\lambda^{\prime}_{ijk}|^2
\bigg[
      -2 h_{u_L}\hat{C}_{24}\left( 0,m_{\tilde{u}_{jL}},m_{\tilde{u}_{jL}}
                            \right) \cr
&   & \qquad\qquad\qquad
      +  h_{d_R}\left\{ (d-2)\hat{C}_{24}\left(m_{\tilde{u}_{jL}},0,0
                                         \right) 
                      - m_Z^2 \hat{C}_{23}\left(m_{\tilde{u}_{jL}},0,0
                                          \right)
                 \right\} \cr
&   & \qquad\qquad\qquad
       +  h_{e_L} B_1\left( 0,m_{\tilde{u}_{jL}} 
                     \right)
      \bigg]  \cr
\delta h_{ijk}^{(u)}
& \equiv & - N_C |\lambda^{\prime}_{ijk}|^2
\bigg[ 2 h_{d_R}
        \hat{C}_{24}\left( m_{u_j},m_{\tilde{d}_{kR}},m_{\tilde{d}_{kR}}
                    \right) \cr
&   &\qquad\qquad\qquad
       - h_{u_L}
        \left\{ (d-2)\hat{C}_{24}\left(m_{\tilde{d}_{kR}},m_{u_j},m_{u_j}
                                \right) 
            - m_Z^2 \hat{C}_{23}\left(m_{\tilde{d}_{kR}},m_{u_j},m_{u_j}
                                \right) 
        \right\}  \cr 
&   &\qquad\qquad\qquad
       + h_{u_R} m_{u_j}^2  
        \hat{C}_0\left(m_{\tilde{d}_{kL}},m_{u_j},m_{u_j}
                 \right) \cr
&   &\qquad\qquad\qquad
       + h_{e_L} B_1\left( m_{u_j},m_{\tilde{d}_{kR}}
                    \right)
\bigg]
\end{eqnarray}
\narrowtext
Again, the dependence on the external momenta has been suppressed.
The corrections which depend on the down--type quark have been grouped
together in $\delta h_{ijk}^{(d)}$ and those that depend on the
up--type quark in $\delta h_{ijk}^{(u)}$.
These combinations are separately finite.
For simplicity, we again evaluate these shifts for a common squark mass of 
$m_{\tilde{q}} = 100$~GeV.

\begin{itemize}
\item  
We begin with the top quark dependent contribution.
For $m_t = 175$~GeV, $m_Z = 92$~GeV, and $m_{\tilde{q}} = 100$~GeV,
we find 
\begin{equation}
\delta h_{i3k}^{(u)} = 0.63\%\,| \lambda^{\prime}_{i3k} |^2.
\end{equation}
This is well approximated by the leading $m_Z = 0$
piece of the expansion in the $Z$ mass:
\begin{equation}
\delta h_{ijk}^{(u)}
\approx - \frac{ N_C }{ 2(4\pi)^2 } | \lambda^{\prime}_{i3k} |^2\, F(x)
\end{equation}
where
\begin{equation}
F(x)= \frac{x}{1-x} \left( 1+ \frac{ 1 }{ 1-x }\ln x
                    \right)
\label{eq:F}
\end{equation}
and $x = m_t^2/m_{\tilde{q}}^2$.
For $m_{\tilde{q}}=100$~GeV, this gives
$0.65\%\,|\lambda^{\prime}_{i3k}|^2$.
The subleading terms from the individual diagrams contributing to
$\delta h_{i3k}^{(u)}$ are:
\begin{eqnarray}
(3b) & : & -\frac{ N_C\,h_{d_R} }{ (4\pi)^2 } |\lambda^{\prime}_{i3k}|^2\, 
            \frac{ m_Z^2 }{ 2 m_{\tilde{q}}^2 } 
            \,f\!\left( \frac{ 1 }{ x } \right)   \cr
(3c) & : & \phantom{+}
            \frac{ N_C\,h_{u_L} }{ (4\pi)^2 } |\lambda^{\prime}_{i3k}|^2\,
            \frac{ m_Z^2 }{ m_t^2 }\,f(x) \cr
(3e) & : & -\frac{ N_C\,h_{u_R} }{ (4\pi)^2 } |\lambda^{\prime}_{i3k}|^2\,
            \frac{ m_Z^2 }{ m_t^2 }\,g(x)
\end{eqnarray}
where 
\begin{eqnarray}
f(x) & \equiv & -\frac{1}{18} \frac{1}{(1-x)^4}
      \left[ 2 x^4 - 9 x^3 +18 x^2 - 11 x - 6 x \ln{x} \right]  
\\
g(x) & \equiv & \phantom{+}\frac{1}{12} \frac{1}{(1-x)^4}
      \left[   x^4 - 6 x^3 + 3 x^2 +  2 x + 6 x^2 \ln{x} \right]
\end{eqnarray}
The total subleading contribution for $m_{\tilde{q}} = 100$~GeV is
$-0.03\%\,|\lambda'_{ijk}|^2$.

In Ref.~\cite{Bhattacharyya:1995pr}, the
leading and subleading contributions of diagrams (3b) and (3c)
are shown\footnote{{\em cf} Eq.~9 of Ref.~\cite{Bhattacharyya:1995pr}.} 
but the subleading contribution of diagram (3e) appears to have been 
omitted. We also disagree with the expression for (3b) in 
Ref.~\cite{Bhattacharyya:1995pr} by a factor of $\frac{1}{2}$.  
However, the numerical impact is negligible.

\item
The corrections involving massless quark loops vanish in the
limit $m_Z \rightarrow 0$ 
and give numerically small contributions.
For the massless up--type quarks ($u_1 = u$ and $u_2 = c$) we find:
\begin{equation}
\delta h_{ijk}^{(u)} = -0.02\%\,|\lambda'_{ijk}|^2.
\end{equation}
The leading order term in $m_Z^2/m_{\tilde{q}}^2$ is
\begin{equation}
\delta h_{ijk}^{(u)}
\approx -\frac{ N_C }{ (4\pi)^2 }|\lambda'_{ijk}|^2
    \left[ h_{u_L} \frac{ m_Z^2 }{ 9m_{\tilde{q}}^2 }
           \left( 1 - 3\ln\frac{ m_Z^2 }{ m_{\tilde{q}}^2 }
           \right)
         - h_{d_R} \frac{ m_Z^2 }{ 18m_{\tilde{q}}^2 }
    \right].
\end{equation}
For $m_{\tilde{q}} = 100$~GeV, this gives $-0.01\%\,|\lambda'_{ijk}|^2$.

\item
The massless down--type quark dependent correction is:
\begin{equation}
\delta h_{ijk}^{(d)} = -0.06\%\,|\lambda'_{ijk}|^2
\end{equation}
The leading order term in $m_Z^2/m_{\tilde{q}}^2$ is
\begin{equation}
\delta h_{ijk}^{(d)}
\approx -\frac{ N_C }{ (4\pi)^2 }|\lambda'_{ijk}|^2
\left[ h_{u_L} \frac{ m_Z^2 }{ 18 m^2_{\tilde{q}} }
     - h_{d_R} \frac{ m_Z^2 }{ 9 m^2_{{\tilde q}} }
               \left( 1 - 3 \ln\frac{ m_Z^2 }{ m^2_{{\tilde q}} }
               \right)
\right] 
\end{equation}
For $m_{\tilde{q}} = 100$~GeV, this gives $-0.09\%\,|\lambda'_{ijk}|^2$.

\end{itemize}

Combining everything together, and summing over
the generation indices $j$ and $k$, the shift of the $i$--th generation
lepton coupling to the $Z$ due to R--violating interactions is:
\begin{eqnarray}
\delta h_i^{\not R}
& = & \sum_{j,k}\left[ \delta h_{ijk}^{(u)} + \delta h_{ijk}^{(d)}
            \right] \cr
& = & 0.61\% \sum_k |\lambda'_{i3k}|^2
    - 0.08\%\sum_k |\lambda'_{i2k}|^2
    - 0.08\%\sum_k |\lambda'_{i1k}|^2 \cr
& \approx & 0.61\%\sum_k |\lambda'_{i3k}|^2
\label{eq:deltah}
\end{eqnarray} 
where we drop the subleading terms.
(This is equivalent to keeping only the diagrams involving the
top and the stop.).

\section{Fits and Numerical Analyses}

In order to place limits on the $\lambda'_{ijk}$ couplings
from $Z$ decay, we need to know how the observables at LEP and SLD
will be affected by the shifts $\delta h_i^{\not R}$ in the 
left--handed coupling of $e_{i}$ to the $Z$, as well as
by other R--conserving vertex and oblique corrections.

The relevant observables are
\begin{equation}
R_\ell = \frac{ \Gamma(Z\rightarrow {\rm hadrons}) }
              { \Gamma(Z\rightarrow  \ell^+\ell^-) }
       = \frac{ N_C \sum_{q=u,d,s,c,b} ( h_{q_L}^2 + h_{q_R}^2 ) }
              { (h_{\ell_L}^2 + h_{\ell_R}^2) }
\end{equation}
and
\begin{equation}
A_\ell = \frac{ h_{\ell_L}^2 - h_{\ell_R}^2 }
              { h_{\ell_L}^2 + h_{\ell_R}^2 }
\end{equation}
as well as
\begin{equation}
A_{\rm FB}(\ell)
= \frac{3}{4} A_e A_\ell
\end{equation}
where $\ell=e$, $\mu$, $\tau$.
The shift in $R_\ell$ due to shifts in the coupling constants is
\begin{eqnarray}
\frac{ \delta R_\ell }{ R_\ell }
& = & \frac{ \delta \Gamma_{\rm had} }{ \Gamma_{\rm had} }
    - \frac{ 2h_{\ell_L}\delta h_{\ell_L} + 2h_{\ell_R}\delta h_{\ell_R} }
           { h_{\ell_L}^2 + h_{\ell_R}^2 } \cr
& = & \Delta_R
    - \frac{ 2h_{\ell_L} }
           { h_{\ell_L}^2 + h_{\ell_R}^2 } \delta h_\ell^{\not R}\cr
& = & \Delta_R + 4.3\; \delta h_\ell^{\not R} \rule[0mm]{0mm}{6mm}
\end{eqnarray}
where we have subsumed all the hadronic corrections and
the lepton flavor independent oblique and
vertex corrections into a single parameter $\Delta_R$,
and the coefficient of $\delta h_\ell^{\not R}$ is calculated
for the value $\sin^2\theta_W = 0.2315$.

We note that the $\lambda'$ couplings we are trying to constrain
also contribute to $\Delta_R$ through $\Gamma_{\rm had}$ since
they modify the $Zq\bar{q}$ vertices\footnote{%
The flavor dependence of R--violating corrections to the
$Zq\bar{q}$ vertices can be used to constrain $\lambda'$ and
$\lambda''$ by looking at purely hadronic observables 
\cite{Lebedev:1999ze}.}
as well as the $Z\ell\bar{\ell}$ vertices.
However, since $\Delta_R$ also subsumes highly {\it model dependent}
corrections from the R--conserving sector, it can be considered
an independent parameter from $\delta h_\ell^{\not R}$ in our fit.


Similarly, the shift in $A_\ell$ is given by
\begin{eqnarray}
\frac{ \delta A_\ell }{ A_\ell }
& = & \frac{ 4 h_{\ell_L} h_{\ell_R}^2 \delta h_{\ell_L}
           - 4 h_{\ell_L}^2 h_{\ell_R} \delta h_{\ell_R}
           }
           { h_{\ell_L}^4 - h_{\ell_R}^4 } \cr
& = & \Delta_A
    + \frac{ 4 h_{\ell_L} h_{\ell_R}^2
           }
           { h_{\ell_L}^4 - h_{\ell_R}^4 } \delta h_{\ell}^{\not R} \cr
& = & \Delta_A -25\;\delta h_{\ell}^{\not R} \rule[0mm]{0mm}{6mm}
\end{eqnarray}
where we have subsumed all the lepton flavor independent oblique and
vertex corrections into a single parameter $\Delta_A$,
and the coefficient of $\delta h_\ell^{\not R}$ is again calculated
for the value $\sin^2\theta_W = 0.2315$.
We do not need to introduce another flavor independent parameter for
$A_{\rm FB}(\ell)$ since
\begin{equation}
\frac{ \delta A_{\rm FB}(\ell) }{ A_{\rm FB}(\ell) }
= \frac{ \delta A_e }{ A_e } + \frac{ \delta A_\ell }{ A_\ell }
= 2\Delta_A - 25\;\delta h_{e}^{\not R} - 25\;\delta h_{\ell}^{\not R}.
\end{equation}
Therefore, we can express all corrections from both
R--breaking and R--conserving interactions in terms of just
5 parameters: $\Delta_R$, $\Delta_A$, and $\delta h_\ell^{\not R}$,
$\ell = e, \mu, \tau$.   A five parameter fit cannot be conducted,
however, since a change in any one parameter can always be absorbed
into the other four.
We therefore define
\begin{eqnarray}
\Delta_{R_e} & \equiv & \Delta_R + 4.3\;\delta h_e^{\not R}, \cr
\Delta_{A_e} & \equiv & \Delta_A - 25\;\delta h_e^{\not R}, \cr
\delta_{\mu e}  & \equiv & \delta h_{\mu }^{\not R} - \delta h_e^{\not R}, \cr
\delta_{\tau e} & \equiv & \delta h_{\tau}^{\not R} - \delta h_e^{\not R}.
\end{eqnarray}
and perform a four parameter fit instead.
Note that the parameters
\begin{eqnarray}
\delta_{\mu  e}
& = & 0.61\%\,\left\{ \sum_k|\lambda'_{23k}|^2
                    - \sum_k|\lambda'_{13k}|^2
              \right\} \cr
\delta_{\tau e}
& = & 0.61\%\,\left\{ \sum_k|\lambda'_{33k}|^2
                    - \sum_k|\lambda'_{13k}|^2
              \right\}
\label{eq:deltas}
\end{eqnarray}
are measures of lepton universality violation.
The dependence of all the observables we use in our fit to
the four fit parameters is:
\begin{eqnarray}
\frac{ \delta R_e    }{ R_e    } & = & \Delta_{R_e} \cr
\frac{ \delta R_\mu  }{ R_\mu  } & = & \Delta_{R_e} + 4.3\;\delta_{\mu  e} \cr
\frac{ \delta R_\tau }{ R_\tau } & = & \Delta_{R_e} + 4.3\;\delta_{\tau e} \cr
\frac{ \delta A_e    }{ A_e    } & = & \Delta_{A_e} \cr
\frac{ \delta A_\mu  }{ A_\mu  } & = & \Delta_{A_e} - 25\;\delta_{\mu  e} \cr
\frac{ \delta A_\tau }{ A_\tau } & = & \Delta_{A_e} - 25\;\delta_{\tau e} \cr
\frac{ \delta A_{\rm FB}(e   ) }{ A_{\rm FB}(e   ) }
                                 & = & 2 \Delta_{A_e} \cr
\frac{ \delta A_{\rm FB}(\mu ) }{ A_{\rm FB}(\mu ) }
                                 & = & 2 \Delta_{A_e} - 25\;\delta_{\mu  e} \cr
\frac{ \delta A_{\rm FB}(\tau) }{ A_{\rm FB}(\tau) }
                                 & = & 2 \Delta_{A_e} - 25\;\delta_{\tau e}
\label{eq:FITCOEFFS} 
\end{eqnarray}

In table~\ref{LEP-SLD-DATA}, we show the most recent data of
these observables from Refs.~\cite{LEP:99}, \cite{SLD:99},
and \cite{BRAU:99}.
The Standard Model predictions
were calculated by ZFITTER v6.21 \cite{ZFITTER:99} with standard
flag settings for the
input values of $m_t = 174.3$~GeV \protect\cite{TOPMASS:99},
$m_H = 300$~GeV, and $\alpha_s(m_Z) = 0.120$.
The limits on lepton universality violation is insensitive to 
the choice of the Higgs mass since Higgs couplings within the
Standard Model do not violate lepton universality by any appreciable
amount.
The correlation matrix of the LEP $Z$ lineshape data is shown in
table~\ref{LEPcorrelations}.

\begin{figure}[p]
\centering
\unitlength=1cm
\begin{picture}(12,10)
\unitlength=1mm
\put(82,80){$R_\mu$}
\put(93,60){$R_\tau$}
\put(48,62){$A_\tau(\mathrm{LEP})$}
\epsfbox[15 50 340 330]{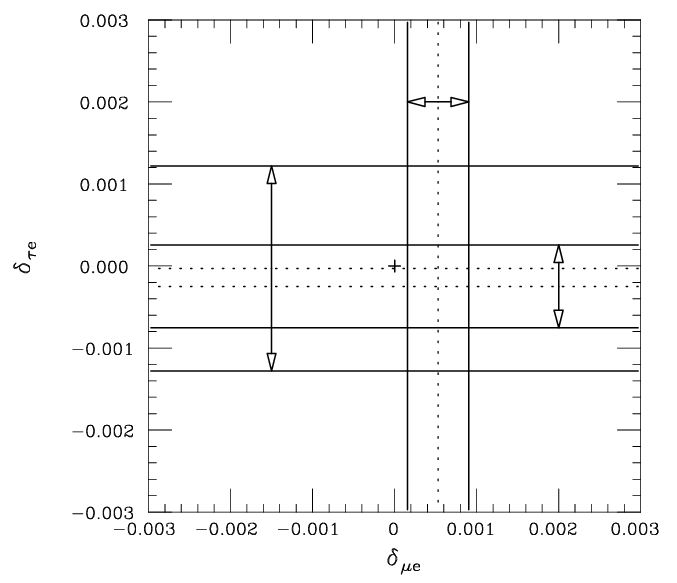}
\end{picture}
\caption{1$\sigma$ constraints on lepton universality violation.}
\label{Bands1}


\centering
\unitlength=1cm
\begin{picture}(12,10)
\unitlength=1mm
\put(86,63){$A_\tau(\mathrm{SLD})$}
\put(78,24){$A_\mu(\mathrm{SLD})$}
\put(33,29){$A_\mathrm{FB}^\tau(\mathrm{LEP})$}
\put(36,74){$A_\mathrm{FB}^\mu(\mathrm{LEP})$}
\epsfbox[15 50 340 330]{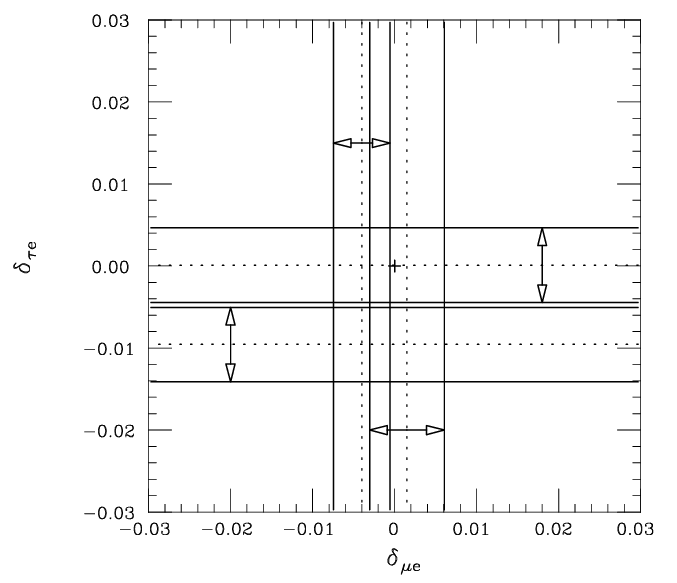}
\end{picture}
\caption{1$\sigma$ constraints on lepton universality violation.
Note the larger scale with respect to the previous figure.}
\label{Bands2}
\end{figure}

\begin{figure}[p]
\centering
\unitlength=1cm
\begin{picture}(12,10)
\unitlength=1mm
\put(80,73){All data}
\put(65,19){lineshape data only}
\put(30,62){asymmetries}
\put(37,58){only}
\epsfbox[15 50 340 330]{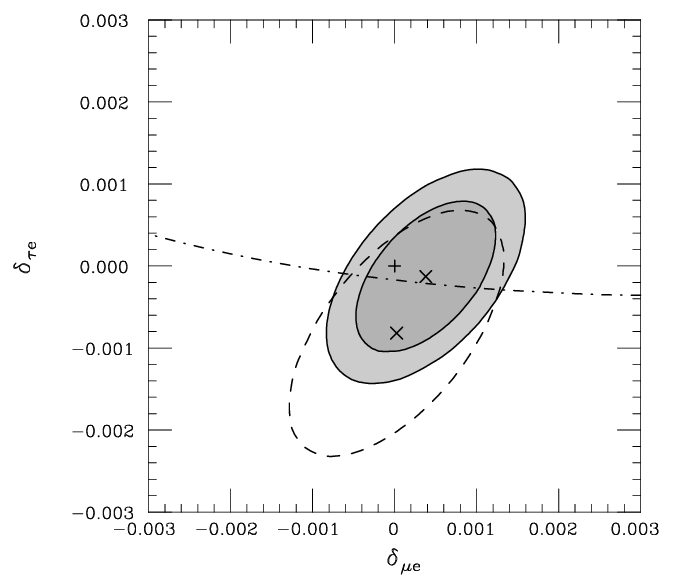}
\end{picture}
\caption{Confidence contours from all data (68\% and 90\%, in gray),
the $Z$ line--shape data only (90\%, dashed line),
and the asymmetry data only (90\%, dot-dashed line).}
\label{Contours1}


\centering
\unitlength=1cm
\begin{picture}(12,10)
\unitlength=1mm
\put(40,40){lineshape data only}
\put(63,60){All data}
\put(70,78){asymmetries only}
\epsfbox[15 50 340 330]{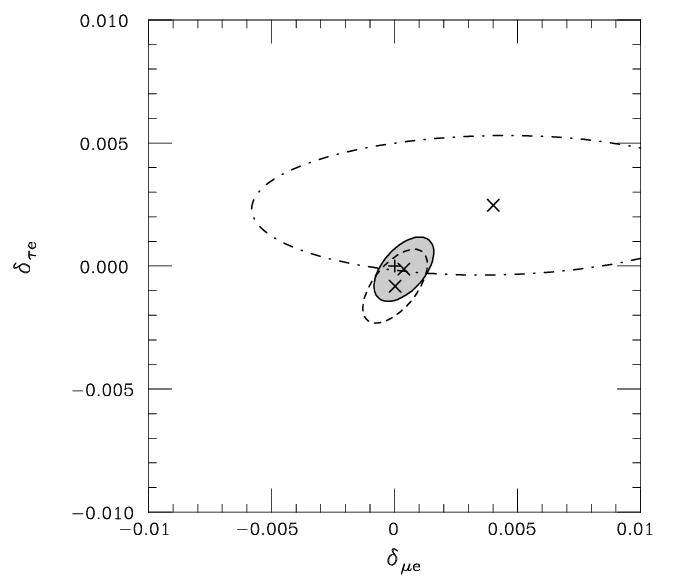}
\end{picture}
\caption{90\% confidence contours from all data (gray),
$Z$ line--shape data only (dashed line), and
asymmetry data only (dot-dashed line). 
Note the different scale with respect to the other figures.
}
\label{Contours2}
\end{figure}

The result of the four parameter fit to all the data in
table~\ref{LEP-SLD-DATA} is
\begin{eqnarray}
\delta_{\mu e}  & = & \phantom{-}0.00038 \pm 0.00056 \cr
\delta_{\tau e} & = & -0.00013 \pm 0.00061\cr
\Delta_{A_e}    & = & \phantom{-}0.052\phantom{0} \pm 0.012 \cr
\Delta_{R_e}    & = & \phantom{-}0.0007 \pm 0.0020
\end{eqnarray}
with the correlation matrix shown in table~\ref{FitCorrelations}.
The quality of the fit was $\chi^2 = 8.3/(12-4)$.
In figures \ref{Bands1} and \ref{Bands2} we show the 1$\sigma$ constraints
placed on $\delta_{\mu e}$ and $\delta_{\tau e}$ in the
$\Delta_{A_e} = \Delta_{R_e} = 0$ plane by each observable.
It is seen that the strongest constraints come from
$R_\mu$, $R_\tau$, and $A_\tau$ from the $\tau$ polarization 
measurement at LEP.
In figure~\ref{Contours1}, we show the 68\% and 90\% confidence
contours on the $\delta_{\mu e}$--$\delta_{\tau e}$ plane.

The limits on $\delta_{\mu e}$ and $\delta_{\tau e}$ translate
into limits on the R--breaking couplings:
\begin{eqnarray}
\sum_k |\lambda'_{23k}|^2 - \sum_k |\lambda'_{13k}|^2 
& = & \phantom{-}0.062 \pm 0.093 \cr
\sum_k |\lambda'_{33k}|^2 - \sum_k |\lambda'_{13k}|^2
& = & -0.02\phantom{0}  \pm 0.10 \cr
\sum_k |\lambda'_{33k}|^2 - \sum_k |\lambda'_{23k}|^2
& = & -0.083 \pm 0.093
\end{eqnarray}
If we neglect the $13k$ terms since they are already constrained
to be small (recall that Eq.~(\ref{eq:1jkconstraint}) shows
the $2\sigma$ upper bound), 
we obtain the following 1$\sigma$ (2$\sigma$) upper bounds 
for the $23k$ and $33k$ terms:
\begin{eqnarray}
\sum_k |\lambda'_{23k}|^2 & \le & 0.16\;(0.25)\cr
\sum_k |\lambda'_{33k}|^2 & \le & 0.08\;(0.18)
\end{eqnarray}
or
\begin{eqnarray}
|\lambda'_{23k}| & \le & 0.40\;(0.50) \cr
|\lambda'_{33k}| & \le & 0.28\;(0.42)
\end{eqnarray}
The limit on $\lambda'_{23k}$ should be interpreted as a limit
on $\lambda'_{232}$ since $\lambda'_{231}$ and $\lambda'_{233}$
are already fairly well constrained by other
experiments \cite{Allanach:1999ic}.
If any of the $13k$--terms (in particular, $\lambda'_{132}$ with
a 2$\sigma$ upper bound of $0.28$ \cite{Allanach:1999ic})
are non--zero, these limits will be weakened.

It is interesting to note that 
since the measured value of $R_\tau$ is smaller than
the measured value of $R_e$, this pair prefers a negative
value of $\delta_{\tau e}$ ({\it cf.} Eq.~\ref{eq:FITCOEFFS}).
The same can be said of the pair $A_{\rm FB}(e)$ and $A_{\rm FB}(\tau)$.
On the other hand,
the measured values of $A_e$ (LEP and SLD) and $A_{\rm LR}$
are all larger than the measured values of $A_\tau$ (LEP and SLD)
so these observables prefer a positive value of $\delta_{\tau e}$.
(This is not apparent in figures~\ref{Bands1} and \ref{Bands2}
since they show the constraints in the $\Delta_{A_e} = 0$ plane.)
Due to this conflict, the central value of $\delta_{\tau e}$
preferred by the global fit is virtually zero which satisfies neither
the $R$'s nor the $A$'s.
In fact, $A_\tau$ from LEP, with its smaller fractional error,
actually accounts for 2.8 out of 6.8 of the $\chi^2$ of the fit.

If we perform our fit on the six $Z$ line--shape
parameters only, the result is
\begin{eqnarray}
\delta_{\mu e}  & = & \phantom{-}0.00002 \pm 0.00061 \cr
\delta_{\tau e} & = & -0.00082 \pm 0.00070\cr
\Delta_{A_e}    & = & \phantom{-}0.055\phantom{0} \pm 0.033 \cr
\Delta_{R_e}    & = & \phantom{-}0.0022 \pm 0.0022
\end{eqnarray}
with $\chi^2 = 1.9/(6-4)$, and the correlation matrix is shown
in table~\ref{RonlyCorrelations}.
The 90\% confidence contour in the $\delta_{\mu e}$--$\delta_{\tau e}$
plane is shown in figures~\ref{Contours1} and \ref{Contours2}.
This translates into
\begin{eqnarray}
\sum_k |\lambda'_{23k}|^2 - \sum_k |\lambda'_{13k}|^2 
& = & \phantom{-}0.00 \pm 0.10 \cr
\sum_k |\lambda'_{33k}|^2 - \sum_k |\lambda'_{13k}|^2
& = & -0.13\phantom{0}  \pm 0.11 \cr
\sum_k |\lambda'_{33k}|^2 - \sum_k |\lambda'_{23k}|^2
& = & -0.138 \pm 0.097
\end{eqnarray}
Again, neglecting the $13k$ term, we obtain the following
1$\sigma$ (2$\sigma$) limits:
\begin{eqnarray}
\sum_k |\lambda'_{23k}|^2 & \le & \phantom{-}0.10\;(0.20)\cr
\sum_k |\lambda'_{33k}|^2 & \le & -0.02\;(0.09)
\end{eqnarray}
or
\begin{eqnarray}
|\lambda'_{23k}| & \le & 0.37\;(0.49) \cr
|\lambda'_{33k}| & \le & \phantom{0.37}\;(0.30)
\label{eq:Ronly}
\end{eqnarray}
The negative central value for $\delta_{\tau e}$
leads to a reduced upper bound for $|\lambda'_{33k}|$.

If we perform our fit on the two $\tau$--polarization observables
and the four SLD observables only, the result is:
\begin{eqnarray}
\delta_{\mu e}  & = & 0.0040 \pm 0.0046 \cr
\delta_{\tau e} & = & 0.0025 \pm 0.0013\cr
\Delta_{A_e}    & = & 0.062 \pm 0.013
\end{eqnarray}
with $\chi^2 = 0.84/(6-3)$, and the correlation matrix is shown
in table~\ref{AonlyCorrelations}.
The 90\% confidence contour in the $\delta_{\mu e}$--$\delta_{\tau e}$
plane for this case is also shown in 
figures \ref{Contours1} and \ref{Contours2}.
This translates into
\begin{eqnarray}
\sum_k |\lambda'_{23k}|^2 - \sum_k |\lambda'_{13k}|^2 
& = & \phantom{-}0.66 \pm 0.75 \cr
\sum_k |\lambda'_{33k}|^2 - \sum_k |\lambda'_{13k}|^2
& = & \phantom{-}0.41 \pm 0.22 \cr
\sum_k |\lambda'_{33k}|^2 - \sum_k |\lambda'_{23k}|^2
& = & -0.25 \pm 0.77
\end{eqnarray}
Neglecting the $13k$ term, we obtain the following
1$\sigma$ (2$\sigma$) limits:
\begin{eqnarray}
\sum_k |\lambda'_{23k}|^2 & \le & 1.4\phantom{0}\;(2.2)\cr
\sum_k |\lambda'_{33k}|^2 & \le & 0.62\;(0.85)
\end{eqnarray}
or
\begin{eqnarray}
|\lambda'_{23k}| & \le & 1.2\phantom{0}\;(1.5) \cr
|\lambda'_{33k}| & \le & 0.79\;(0.92)
\end{eqnarray}
This time, the upper bounds are considerably larger.

This shows that had we used only the $Z$ line--shape
variables, which has been the case in previous analyses by
other authors \cite{Bhattacharyya:1995pr,YANG:99}, or
only the leptonic asymmetries, we would have reached drastically
different conclusions concerning the limits on the R--breaking parameters.
Only through a {\it global} analysis were we able to
constrain the parameters in a consistent way.

\section{Summary and Conclusions}

We find that flavor-{\it conserving} leptonic $Z$ and $W$ decays
can be used to place significant constraints on 
the size of R--parity violating $\lambda^\prime$ couplings.  
Current
bounds from lepton universality violation in leptonic $Z$ decays from 
combined LEP/SLD data are
\begin{eqnarray}
|\lambda'_{23k}| & \le & 0.40\;(0.50) \cr
|\lambda'_{33k}| & \le & 0.28\;(0.42)
\end{eqnarray}
at the 1$\sigma$ (2$\sigma$) level,
assuming a common squark mass of $m_{\tilde{q}}=100$~GeV and the
suppression of $\lambda'_{13k}$ couplings.
For larger (common) squark masses the above bounds should be interpreted
as bounds on $|\lambda'| \times \sqrt{F(x)/F(x_0)}$, where $F(x)$ is defined
in Eq. \ref{eq:F} and $x_0=\frac{m_t^2}{(100 {\rm GeV})^2}.$

Numerically, our numbers are not a significant improvement
over those cited in Ref.~\cite{Allanach:1999ic}.
However, the methods used to derive previous limits
\cite{Bhattacharyya:1995pr,YANG:99} were
intrinsically flawed in that (1) R--conserving
effects were not properly taken into account and
(2) R--breaking effects on only the leptonic widths of the 
$Z$ were considered.
Indeed, had we also considered only the leptonic widths, our
limits would have been those of Eq.~(\ref{eq:Ronly}).
The analysis of this paper avoids these problems by
focussing on {\it lepton universality violation} and
performing a {\it global} fit on all
LEP/SLD observables.
The R--conserving effects are taken into account by
parametrizing and fitting them to the data also.
(Similar methods have been used in Ref.~\cite{LOINAZ:99}
to constrain flavor specific vertex corrections while
taking into account the flavor universal oblique corrections.)

Current bounds on lepton universality in leptonic $W$ decays 
provides the constraint
\begin{equation}
|\lambda^{\prime}_{33k}| \leq 1.7 \; (2.4)
\end{equation}
at the 1$\sigma$ (2$\sigma$) level.  
While not currently competitive with the $Z$ decay bounds, 
the Fermilab results are complementary independent measurements, 
and they can be expected to improve dramatically at Tevatron Run~II.

If the error on the LEP/SLD observables continue to shrink with
the current central values, then eventually
the region allowed by the line--shape variables and the asymmetries
will fail to overlap in figure~\ref{Contours2}.
In such a situation, not only the SM but the MSSM with R--parity
violating couplings would be ruled out.
In fact, no theory which introduces lepton universality violation in
only the left--handed couplings would be viable.

Currently, the LEP and SLD observable provide the best
limits on the $\lambda'_{33k}$ couplings.  
However, one can potentially place a limit
on the $\lambda'$ couplings by looking at invisible decays
of the $\Upsilon$ and $J/\Psi$ resonances 
at the $B$ and $\tau$--charm factories \cite{CHANG:98}.
The current bounds on $\lambda'_{i33}$ imply that the correction
to the invisible width of the $\Upsilon$ resonance can be
as large as 30\%.
A rough estimate shows that if the $\Upsilon$ invisible width
is found to agree with the Standard Model prediction
with 5\% accuracy, the $\lambda'$ coupling would be constrained to be
$|\lambda'_{i33}| \leq 0.16$ at the 2$\sigma$ level.  In addition, 
constraints on $|\lambda'_{333}|$ will be available from forthcoming Tevatron 
studies of the decay $t \rightarrow \tau b$ \cite{Han:1999qs}.


\acknowledgements

We thank Morris Swartz and Robert Clare for providing us with the
latest LEPEWWG data including the correlation matrices.
Helpful communications with Gautam Bhattacharyya,
James E. Brau, Herbi Dreiner, Apostolos Pilaftsis, Franco Rimondi, and 
Peter Rowson are also gratefully acknowledged.
We thank the hospitality of the Fermilab Particle 
Theory Group, where part of this work was conducted under the 
auspices of the Summer Visitors' Program.
This work was supported in part (O.L. and W.L.) by the 
U.~S. Department of Energy, grant DE-FG05-92-ER40709, Task A.


\onecolumn
\widetext

\appendix
\section*{Feynman Integrals}

\newcommand{\pole}{\Delta_\epsilon}

\setlength{\jot}{2mm}
\setlength{\abovedisplayskip}{4mm plus 1mm minus 1mm}
\setlength{\belowdisplayskip}{4mm plus 1mm minus 1mm}

Here we make explicit our notation for the scalar and tensor integrals 
that appear in the calculation.
The definitions of the integrals are similar to those of
Ref.~\cite{'tHooft:1979xw}, but there are some small
alterations made to take advantage of symmetries of the problem.  
The hat on the tensor integrals serves as a reminder of these differences.

\subsection{Scalar Integrals}

\noindent
We define the functions $B_0$ and $\hat{C}_0$ by:
\begin{eqnarray}
B_0\,[\,p^2;m_1,m_2\,] & \equiv & 
i\,\mu^{4-d}\int\!\frac{ d^d k }{ (2\pi)^d }\, 
\frac{ 1 }{ (k^2 - m_1^2)\,[\,(k+p)^2 - m_2^2\,] } \\
\hat{C}_0\,[\,p^2,q^2,(p-q)^2;m_1,m_2,m_3\,] & \equiv &
i\int\!\frac{ d^4 k }{ (2\pi)^4 }\, 
\frac{ 1 }
     { ( k^2 - m_1^2 )\,[\,(k+p)^2 - m_2^2\,]\,[\,(k+q)^2 - m_3^2\,] }
\end{eqnarray}
The general form of $B_0$ is given by
\begin{displaymath}
B_0\,[\,p^2;m_1,m_2\,] 
= \frac{-1}{(4\pi)^2}
      \left[ \pole 
             - \dfrac{ m_1^2\ln(m_1^2/\mu^2) - m_2^2\ln(m_2^2/\mu^2) }
                     { m_1^2 - m_2^2 }
             + 1 + F(p^2;m_1,m_2)
      \right]
\end{displaymath}
where $\pole = \dfrac{2}{4-d} - \gamma_E + \ln{4\pi}$, and \cite{HOLLIK:90}
\begin{equation}
F(p^2;m_1,m_2) = 1
  + \frac{1}{2} \left( \frac{\Sigma}{\Delta} - \Delta 
                \right) \ln\left( \frac{m_1^2}{m_2^2}
                           \right)
  - \frac{1}{2}\sqrt{ 1 - 2\Sigma + \Delta^2 } 
  \,\ln\left( \dfrac{ 1 - \Sigma + \sqrt{ 1 - 2\Sigma + \Delta^2 } }
                    { 1 - \Sigma - \sqrt{ 1 - 2\Sigma + \Delta^2 } }
       \right)
\end{equation}
with
\begin{equation}
\Sigma \equiv \dfrac{m_1^2 + m_2^2}{p^2},\qquad
\Delta \equiv \dfrac{m_1^2 - m_2^2}{p^2}.
\label{SigmaDelta}
\end{equation}
The function $F(p^2;m_1,m_2)$ is well--behaved in the limit
$p^2\rightarrow 0$.   For small $p^2\ll m_1^2, m_2^2$, the behavior is:
\begin{eqnarray}
F(p^2;m_1,m_2)
& = & \frac{ p^2 }{ 2(m_1^2-m_2^2)^2 }
      \left[ m_1^2 + m_2^2
           - \frac{2\,m_1^2 m_2^2}{m_1^2 - m_2^2}
             \ln\left( \frac{m_1^2}{m_2^2} \right)
      \right] + \cdots
\label{HollikF}\\
F(p^2;m,m) & = & \frac{p^2}{6m^2} + \cdots
\\
F(p^2;0,m) & = & \frac{p^2}{2m^2} + \cdots
\end{eqnarray}

\noindent
Useful special cases of the $B_0$ function are:
\begin{eqnarray}
B_0\,[\,p^2;m,m\,] & = & 
\frac{-1}{(4\pi)^2}
\left[ \pole - \ln\frac{m^2}{\mu^2} + 2
       -\sqrt{ 1 - \frac{4m^2}{p^2} }\,
        \ln\left( \frac{ \sqrt{ 1 - \frac{4m^2}{p^2} } + 1 }
                       { \sqrt{ 1 - \frac{4m^2}{p^2} } - 1 }
           \right)
\right]
\\
B_0\,[\,p^2;0,m\,] & = &
\frac{-1}{(4\pi)^2}
\left[ \pole - \ln\frac{m^2}{\mu^2} + 2
       -\left( 1-\frac{m^2}{p^2} 
        \right)
        \ln\left( 1-\frac{p^2}{m^2}
           \right)
\right]
\end{eqnarray}

\noindent
The general form of the $\hat{C}_0$ function is fairly complex
and we refer the reader to Ref.~\cite{'tHooft:1979xw}.
It simplifies considerably for the following cases:
\begin{eqnarray}
\hat{C}_0\,[\,0,0,p^2;m,0,0\,] & = &
\frac{ -1 }{ (4\pi)^2 } \frac{ 1 }{ p^2 } 
\left[\,\ln\left(\frac{ p^2 }{ m^2 }
           \right)
        \ln\left( 1 + \frac{ p^2 }{ m^2 }
           \right)
     + {\rm Li}_2\left( -\frac{ p^2 }{ m^2 }
                 \right)\,
\right]
\\
\hat{C}_0\,[\,0,0,p^2;0,m_1,m_2\,] & = & 
\frac{ -1 }{ (4\pi)^2 } \frac{ 1 }{ 4p^2 }
\left[\, \ln^2\left( \frac{ 1 - \Sigma + \sqrt{ 1 - 2\Sigma + \Delta^2 } }
                          { 1 - \Sigma - \sqrt{ 1 - 2\Sigma + \Delta^2 } }
              \right)
       - \ln^2\left( \frac{ m_1^2 }{ m_2^2 }
              \right)\,
\right] 
\end{eqnarray}
where $\Sigma$ and $\Delta$ are defined as in Eq.~(\ref{SigmaDelta}).

The following $\hat{C}_0$ could be expressed in terms 
a sum of dilogarithms, but for our purposes it is
simpler to reduce them to a Feynman parameter integral and either 
perform the integration numerically or, 
if an expansion is needed, to expand the integrand directly and 
then integrate.
\begin{eqnarray}
\hat{C}_0\,[\,0,0,p^2;m,M,0\,]
& = & \frac{ 1 }{ (4\pi)^2 } \int_0^1\!dx\,
      \frac{ 1 }{ p^2 (1-x) + m^2 }\,
      \ln\left[ \frac{ (m^2-M^2) x + M^2 }{ (1-x)(M^2 - x p^2 ) }
         \right] \\
\hat{C}_0\,[\,0,0,p^2;m,M,M\,]
& = & \frac{ 1 }{ (4\pi)^2 } \int_0^1\!dx\, 
      \frac{ 1 }{ p^2 (1-x) + (m^2 - M^2) }\,
      \ln\left[ \frac{ (m^2 - M^2) x + M^2 }{ -p^2 x (1-x) + M^2 } 
         \right] 
\end{eqnarray}

\subsection{Tensor Integrals}

\noindent
Definition and general form of $B_1$:
\begin{eqnarray}
B_\mu\,[\,p;m_1,m_2\,]
& = & i\mu^{4-d} \int\!\frac{ d^dk }{ (2\pi)^d }\,
      \frac{ k_\mu }{ (k^2 - m_1^2)\,[(k+p)^2 - m_2^2] }
\;\equiv\; p_{\mu} B_1\,[\,p^2;m_1,m_2\,]  \\
B_1\,[\,p^2;m_1,m_2\,]
& = & - \frac{1}{2} B_0\,[\,p^2;m_1,m_2\,]
      + \frac{ 1 }{ (4\pi)^2 }
        \left( \frac{ m_1^2 - m_2^2 }{ 2p^2 }
        \right)
        F(p^2;m_1,m_2)
\end{eqnarray}
>From Eq.~(\ref{HollikF}), we find that in the limit $p^2\rightarrow 0$:
\begin{equation}
B_1\,[\,0;m_1,m_2\,]
=  -\frac{1}{2} B_0\,[\,0;m_1,m_2\,]
  + \frac{ 1 }{ (4\pi)^2 }
    \frac{ 1 }{ 4(m_1^2 - m_2^2)}
    \left[ m_1^2 + m_2^2
         - \frac{ 2\,m_1^2 m_2^2 }{ m_1^2 - m_2^2 }
           \ln\left( \frac{ m_1^2 }{ m_2^2 }
              \right)
    \right]
\end{equation}
Other special cases:
\begin{eqnarray}
B_1\,[\,p^2;0,m\,]
& = & \frac{ 1 }{ (4\pi)^2 }\,\frac{ 1 }{ 2 }
      \left[ \pole - \ln\left( \frac{ m^2 }{ \mu^2 } \right)
           + 2 - \frac{ m^2 }{ p^2 }
           - \left( 1 - \frac{ m^2 }{ p^2 } \right)^2 
             \ln\left( 1 - \frac{ p^2 }{ m^2 } \right)
      \right] \\
& \stackrel{p^2\rightarrow 0}{\Longrightarrow} & 
      \frac{ 1 }{ (4\pi)^2 }\,\frac{ 1 }{ 2 }
      \left[ \pole - \ln\left( \frac{ m^2 }{ \mu^2 } \right)
           + \frac{1}{2}
      \right]
\end{eqnarray}
Useful relations among the $B$--functions:
\begin{eqnarray}
0 & = & B_0\,[\,p^2;m_1,m_2\,]
      + B_1\,[\,p^2;m_1,m_2\,]
      + B_1\,[\,p^2;m_2,m_1\,],\\ 
0 & = & (m_1^2-m_2^2)\,B_0\,[\,0;m_1,m_2\,]
      + (m_2^2-m_3^2)\,B_0\,[\,0;m_2,m_3\,]
      + (m_3^2-m_1^2)\,B_0\,[\,0;m_3,m_1\,].
\end{eqnarray}
Definition of the $C$--functions: (Note the difference from the
definitions in Ref.~\cite{'tHooft:1979xw}.)
\begin{eqnarray}
C_\mu\,[\,p,q;m_1,m_2,m_3\,]
& = & i\int\!\frac{ d^4 k }{ (2\pi)^4 }
      \frac{ k_\mu }
           { (k^2 - m_1^2)\,[ (k+p)^2 - m_2^2 ]\,[ (k+q)^2 - m_3^2 ] }
\nonumber\\
& \equiv & p_\mu \hat{C}_{11} + q_\mu \hat{C}_{12}
\\
C_{\mu\nu}[\,p,q;m_1,m_2,m_3\,]
& = & i\mu^{4-d}\int\!\frac{ d^d k }{ (2\pi)^d }
      \frac{ k_\mu k_\nu }
           { (k^2 - m_1^2)\,[ (k+p)^2 - m_2^2 ]\,[ (k+q)^2 - m_3^2 ] }
\nonumber\\
& \equiv & p_\mu p_\nu \hat{C}_{21} + q_\mu q_\nu \hat{C}_{22}
          + ( p_\mu q_\nu + q_\mu p_\nu ) \hat{C}_{23}
          + g_{\mu\nu} \hat{C}_{24} \nonumber
\end{eqnarray}
For the purpose of this paper, we will only need to evaluate these functions
for $p^2 = q^2 = 0$ (we neglect final state fermion masses).
$Q^2 = (p-q)^2 = -2p\cdot q$ will then be the invariant mass squared of the
initial vector boson{\footnote{ We caution the reader that $p$ and $q$ defined here
are different from those appearing in the figures.}}.  For this parameter choice, 
the $C$--functions can be expressed in terms of the $B$--functions
and $\hat{C}_0$ as:
\begin{eqnarray}
\hat{C}_{11}
& = & -\frac{ 1 }{ Q^2 }
       \left\{ B_0\,[\,  0;m_1,m_2\,]
             - B_0\,[\,Q^2;m_2,m_3\,] 
             - (m_1^2-m_3^2)\,
               \hat{C}_0
       \right\}
\\
\hat{C}_{12}
& = & -\frac{ 1 }{ Q^2 }
       \left\{ B_0\,[\,  0;m_1,m_3\,]
             - B_0\,[\,Q^2;m_2,m_3\,] 
             - (m_1^2-m_2^2)\,
               \hat{C}_0
       \right\}
\\
(d-2)\hat{C}_{24}
& = & - B_1\,[\,Q^2;m_2,m_3\,]
      + (m_1^2-m_2^2)\,\hat{C}_{11} + m_1^2\,\hat{C}_0
\\
-Q^2\,\hat{C}_{23} + 2\,\hat{C}_{24}
& = & - B_1\,[\,Q^2;m_2,m_3\,] - (m_1^2-m_3^2)\,\hat{C}_{12} \cr
& = & - B_1\,[\,Q^2;m_3,m_2\,] - (m_1^2-m_2^2)\,\hat{C}_{11}
\end{eqnarray}
We do not list expressions for $\hat{C}_{21}$ nor $\hat{C}_{22}$
since we do not use them in this paper.

\narrowtext

\begin{table}[p]
\begin{center}
\begin{tabular}{|c|c|c|}
Observable & Measured Value & ZFITTER Prediction \\
\hline\hline
\multicolumn{2}{|l|}{\underline{$Z$ lineshape variables}} & \\
$m_Z$                & $91.1872 \pm 0.0021$ GeV & input       \\
$\Gamma_Z$           & $2.4944 \pm 0.0024$ GeV  & unused      \\
$\sigma_{\rm had}^0$ & $41.544 \pm 0.037$ nb    & unused  \\
$R_e$                & $20.803 \pm 0.049$       & $20.739$ \\
$R_\mu$              & $20.786 \pm 0.033$       & $20.739$ \\
$R_\tau$             & $20.764 \pm 0.045$       & $20.786$ \\
$A_{\rm FB}(e   )$   & $0.0145 \pm 0.0024$      & $0.0152$ \\
$A_{\rm FB}(\mu )$   & $0.0167 \pm 0.0013$      & $0.0152$ \\
$A_{\rm FB}(\tau)$   & $0.0188 \pm 0.0017$      & $0.0152$ \\
\hline
\multicolumn{2}{|l|}{\underline{$\tau$ polarization at LEP}} & \\
$A_e$        & $0.1483 \pm 0.0051$      & $0.1423$ \\ 
$A_\tau$     & $0.1424 \pm 0.0044$      & $0.1424$ \\
\hline
\multicolumn{2}{|l|}{\underline{SLD left--right asymmetries}} & \\
$A_{LR}$     & $0.15108 \pm 0.00218$    & $0.1423$ \\
$A_e$        & $0.1558 \pm 0.0064$    & $0.1423$ \\
$A_{\mu}$    & $0.137 \pm 0.016$    & $0.1423$ \\
$A_{\tau}$   & $0.142 \pm 0.016$    & $0.1424$ \\
\end{tabular}
\caption{LEP/SLD observables 
and their Standard Model predictions.
The $Z$ lineshape observables are from Ref.~\protect\cite{LEP:99}.
The rest of the data is from Ref.~\protect\cite{SLD:99} and
\protect\cite{BRAU:99}.
The Standard Model predictions were calculated using ZFITTER v.6.21 
\protect\cite{ZFITTER:99} with $m_t = 174.3$~GeV \protect\cite{TOPMASS:99},
$m_H = 300$~GeV, and $\alpha_s(m_Z) = 0.120$ as input.}
\label{LEP-SLD-DATA}
\end{center}
\end{table}

\medskip

\widetext

\begin{table}[ht]
\begin{center}
\begin{tabular}{|c|ccccccccc|}
& $m_Z$     & $\Gamma_Z$     & $\sigma_{\rm had}^0$
& $R_e$     & $R_\mu$     & $R_\tau$ 
& $A_{\rm FB}(e)$ & $A_{\rm FB}(\mu)$ & $A_{\rm FB}(\tau)$ \\
\hline
$m_Z$ 
& $1.000$            & $-0.008$           & $-0.050$ 
& $\phantom{-}0.073$ & $\phantom{-}0.001$ & $\phantom{-}0.002$ 
& $-0.015$           & $\phantom{-}0.046$ & $\phantom{-}0.034$ \\
$\Gamma_Z$
&                    & $\phantom{-}1.000$ & $-0.284$ 
& $-0.006$           & $\phantom{-}0.008$ & $\phantom{-}0.000$ 
& $-0.002$           & $\phantom{-}0.002$ & $-0.003$           \\
$\sigma_{\rm had}^0$
&                    &                    & $\phantom{-}1.000$ 
& $\phantom{-}0.109$ & $\phantom{-}0.137$ & $\phantom{-}0.100$ 
& $\phantom{-}0.008$ & $\phantom{-}0.001$ & $\phantom{-}0.007$ \\
$R_e$
&                    &                    & 
& $\phantom{-}1.000$ & $\phantom{-}0.070$ & $\phantom{-}0.044$ 
& $-0.356$           & $\phantom{-}0.023$ & $\phantom{-}0.016$ \\
$R_\mu$
&                    &                    & 
&                    & $\phantom{-}1.000$ & $\phantom{-}0.072$ 
& $\phantom{-}0.005$ & $\phantom{-}0.006$ & $\phantom{-}0.004$ \\
$R_\tau$ 
&                    &                    &
&                    &                    & $\phantom{-}1.000$ 
& $\phantom{-}0.003$ & $-0.003$           & $\phantom{-}0.010$ \\
$A_{\rm FB}(e)$ 
&                    &                    &
&                    &                    & 
& $\phantom{-}1.000$ & $-0.026$           & $-0.020$ \\
$A_{\rm FB}(\mu)$ 
&                    &                    & 
&                    &                    & 
&                    & $\phantom{-}1.000$ & $\phantom{-}0.045$ \\
$A_{\rm FB}(\tau)$
&                    &                    & 
&                    &                    & 
&                    &                    & $\phantom{-}1.000$ \\
\end{tabular}
\caption{The correlation of the $Z$ lineshape variables at LEP}
\label{LEPcorrelations}
\end{center}
\end{table}

\narrowtext

\begin{table}[ht]
\begin{center}
\begin{tabular}{|c|cccc|}
& $\delta_{\mu e}$ & $\delta_{\tau e}$ & $\Delta_A$ & $\Delta_R$ \\
\hline\hline
$\delta_{\mu  e}$ & $1.00$ & $0.53$ & $0.22$ & $-0.76$  \\
$\delta_{\tau e}$ &        & $1.00$ & $0.28$ & $-0.63$  \\
$\Delta_A$        &        &        & $1.00$ & $-0.23$ \\
$\Delta_R$        &        &        &         & $\phantom{-}1.00$  \\
\end{tabular}
\caption{The correlation matrix of the fit parameters using all data.}
\label{FitCorrelations}
\end{center}
\end{table}

\narrowtext

\begin{table}[ht]
\begin{center}
\begin{tabular}{|c|cccc|}
& $\delta_{\mu e}$ & $\delta_{\tau e}$ & $\Delta_A$ & $\Delta_R$ \\
\hline\hline
$\delta_{\mu  e}$ & $1.00$ & $0.60$ & $0.32$ & $-0.79$  \\
$\delta_{\tau e}$ &        & $1.00$ & $0.29$ & $-0.69$  \\
$\Delta_A$        &        &        & $1.00$ & $-0.33$  \\
$\Delta_R$        &        &        &        & $\phantom{-}1.00$  \\
\end{tabular}
\caption{The correlation matrix of the fit parameters using the 
$Z$ line--shape data only.}
\label{RonlyCorrelations}
\end{center}
\end{table}

\narrowtext

\begin{table}[ht]
\begin{center}
\begin{tabular}{|c|ccc|}
& $\delta_{\mu e}$ & $\delta_{\tau e}$ & $\Delta_A$  \\
\hline\hline
$\delta_{\mu  e}$ & $1.00$ & $0.05$ & $0.12$ \\
$\delta_{\tau e}$ &        & $1.00$ & $0.41$ \\
$\Delta_A$        &        &        & $1.00$ \\
\end{tabular}
\caption{The correlation matrix of the fit parameters using the 
LEP $\tau$--polarization and SLD leptonic asymmetries only.}
\label{AonlyCorrelations}
\end{center}
\end{table}

\widetext

\narrowtext

\end{document}